\documentclass{pasa}%

\usepackage{amsmath}
\usepackage{epsfig}

\newfont{\myfont}{cmmib10}

\newfont{\myfontsmall}{cmmib8}

\DeclareSymbolFont{cmmi}{OML}{cmm}{m}{it}
\DeclareMathSymbol{v}{\mathalpha}{cmmi}{"76}

\newcommand{\gesim}{\,\raisebox{-0.4ex}{$\stackrel{>}{\scriptstyle\sim}$}\,}
\newcommand{\lesim}{\,\raisebox{-0.4ex}{$\stackrel{<}{\scriptstyle\sim}$}\,}

\def\ga{\gesim}
\def\la{\lesim}

\def\l<{\langle}
\def\r>{\rangle}

\title[Strategies for Slow Transients Surveys]{Optimization of Survey Strategies for Detecting Slow Radio Transients}
\author[Macquart]{Jean-Pierre Macquart$^1$\thanks{email J.Macquart@curtin.edu.au}\\
\affil{$^1$ICRAR/Curtin Institute of Radio Astronomy, GPO Box U1987, Perth, WA 6845, Australia }%
\affil{$^2$ARC Centre of Excellence for All-Sky Astrophysics (CAASTRO)}}%
\jid{PASA}
\doi{10.1017/pas.\the\year.xxx}
\jyear{\the\year}

\begin{document}
\begin{abstract}
We investigate the optimal tradeoff between sensitivity and field of view in surveys for slow radio transients using the event detection rate as the survey metric. This tradeoff bears implications for the design of surveys conducted with upcoming widefield radio interferometers, such as the ASKAP VAST survey and the MeerKAT TRAPUM survey.   We investigate (i) a survey in which the events are distributed homogeneously throughout a volume centred on the Earth, (ii) a survey in which the events are homogeneously distributed, but are only detectable beyond a certain minimum distance, and (iii) a survey in which all the events occur at an identical distance, as is appropriate for a targetted survey of a particular field which subtends $N_{\rm point}$ telescope pointings.  For a survey of fixed duration, $T_{\rm obs}$, we determine the optimal tradeoff between number of telescope pointings, $N$, and integration time per field.  We consider a population in which the event luminosity distribution follows a power law with index $-\alpha$, and $t_{\rm slew}$ is the slewing time between fields or, for a drift scan, the time taken for the telescope drift by one beamwidth.  Several orders of magnitude improvement in detection rate is possible by optimization of the survey parameters.  The optimal value of $N$ for case (i) is $N_{\rm max} \sim T_{\rm obs}/4 t_{\rm slew}$, while for case (iii) we find $N_{\rm max} =  (L_{\rm max}/L_0)^2 [ (3 -\alpha)/2 ]^{2/(\alpha-1)}$, where $L_{\rm max}$ is the maximum luminosity of a transient event and $L_0$ is the minimum luminosity event detectable in an integration of duration $T_{\rm obs}$.   (The instance $N_{\rm max} > N_{\rm point}$ in (iii) implies re-observation of fields over the survey area, except when the duration of transient events exceeds that between re-observations of the same field, where $N_{\rm max}=N_{\rm point}$ applies instead.)
We consider the balance in survey optimization between telescope field of view, $\Omega$, and sensitivity, characterized by the minimum detectable flux density, $S_0$.  For homogeneously distributed events (i), the detection rate scales as $N \Omega S_0^{-3/2}$, while for targetted events (iii) it scales as $N \Omega S_0^{1-\alpha}$.  However, if the targetted survey is optimized for $N$ the event detection rate scales instead as $\Omega S_0^{-2}$.   This analysis bears implications for the assessment of telescope designs: the quantity $\Omega S_0^{-2}$ is often used as the metric of telescope performance in the SKA transients literature, but only under special circumstances is it the metric that optimizes the event detection rate.
\end{abstract}
\maketitle%

\onecolumn
%
%

\section{Introduction}
The advent of widefield and sensitive radio telescopes such as LOFAR, the MWA, MeerKAT, the JVLA and ASKAP has imparted renewed impetus to surveys of the transient radio sky (Bower et al.\,2007; Croft et al.\,2010; Frail et al.\,2012; Mooley et al.\,2013; Murphy 2013).  The most common and straightforward search mode employed on interferometric facilities such as these involves inspection of the data in the imaging domain.  Imaging surveys are restricted to transients on timescales greater than the telescope correlator integration timescale, typically 3-10 seconds, and typically span event durations between several seconds up to days or months.

The domain of astrophysical radio transient events remains poorly characterized.  The large volume of unexplored parameter space that may be occupied by transient events means that even fundamental observables of the whole population, such as the number of transients detectable as a function of flux density or timescale, are poorly known (e.g. Bower et al.\,2007; Frail et al.\,2012).  It is often for this reason that several survey groups espouse a variety of different survey techniques (e.g. VAST), based on the philosophy that various survey optimizations are necessary to cover the diversity of all possible transient astrophysical phenomena (Murphy et al.\,2013).  

The fact that transient astrophysical phenomena may span a large range in timescale, luminosity and frequency need not necessarily preclude the existence of an optimal strategy capable of simultaneously capturing events covering a large range of these parameters.  It is the purpose of this paper to explore the optimization of imaging transients surveys in order to  capture the largest number of events.

There are several important differences between fast and slow transients that dictate how to survey them optimally.   Radio transients can be distinguished as either slow or fast on the basis of their duration relative to the typical integration timescale of the telescope correlator.  It is usually necessary to dedisperse the signals associated with fast transients in order to maximise their signal-to-noise ratio, and the short duration of their emission renders them subject to temporal smearing caused by scattering in interstellar and intergalactic plasmas.  A defining characteristic of slow transients is that they occur on a timescale too slow for either dispersion or scattering to significantly diminish their signal-to-noise ratio; thus these effects are not relevant considerations in the context of slow transients surveys.  In this paper we are concerned with strategies for imaging surveys of slow transients only.

An important consideration is the duration of the events relative to the telescope survey time.  If the interval between two successive observations of a given field is large compared to the typical transient event duration, then the observations are statistically independent, and we may expect to discover a disjoint set of transients between observations.  It is convenient to introduce the formal concept of an event decoherence time as the timescale over which independence applies.  If the interval between the observations is short compared to this timescale, then one expects to redetect essentially the same set of transients discovered in the first observation of that field.  Fast and slow transients may also thus be distinguished on the basis of whether the event decoherence time is short or long compared to the interval that the telescope dwells on a given field.  According to this definition, it is possible to search any given field of view for slow transients down to a given sensitivity limit only infrequently, and to use the intervening time between searches of a given field to survey other fields for further slow transients.  It is this defining characteristic of slow transients that forms the basis of the present optimization analysis.  

A crucial consideration in any transients survey is the balance between sensitivity and field of view.  Is it more advantageous to visit in succession a small number of fields of view with high sensitivity, or a large number of fields of view with accordingly lower sensitivity?  The objective of this paper is to determine the optimal balance between total survey field of view and sensitivity for a survey of slow transients, and to consider how the choice of these parameters alters the event detection rate.

The structure of this paper is as follows.  In section \ref{Decohere} we preface our discussion of event detection rates by making a formal definition of the event decoherence time and compute its value for several likely distributions of event durations.  In section \ref{SingleField} we relate the detection rate of transients for a single field to the underlying luminosity function of the parent transient population.  In section \ref{MultiFields} we then consider the increase in detection rate obtained by subdividing the observation into a survey of $N$ fields.  Our objective is to find the value for $N$ that optimises the balance between the sensitivity reached per field and the total area that may be surveyed in a given time $T_{\rm obs}$ so that the detection rate is maximised.  Section \ref{Discussion} discusses the implications of our results by presenting a set of simple rules for optimizing the survey strategy, and section \ref{Conclusion} presents our conclusions.

\section{Survey Preliminaries and Fundamentals} \label{Decohere}

Our analysis is based on the premise that a survey is optimal if it detects a maximal number of unique transient events.  Survey optimization therefore depends on several fundamental attributes of both the events we wish to detect and of our survey instrument, namely: (1) the event decoherence time,  which is a measure of the interval of time between which successive visits to the same field detect a disjoint set of transient events (see below), (2) the telescope sensitivity, (3) the luminosity distribution function of the population, particularly its slope, and (4) the manner of survey we wish to undertake.  In relation to (4), we consider here two archetypical surveys: (i) those which target events are distributed homogeneously throughout space, and (ii) those which target events at fixed distance, as is appropriate for events embedded in a galaxy or a cluster of galaxies.  This work extends the formalism originally introduced in Macquart (2011) to treat the detection of fast transients.

Another consideration is the effect of interstellar scintillation, which can alter the flux density of any compact object whose radiation propagates through the interstellar medium of our Galaxy.  Most slow transients are not sufficiently compact to exhibit the fast, large amplitude variability caused by diffractive interstellar scintillation, but they are compact enough to undergo refractive scintillation.  Although refractive scintillation is a widespread phenomenon amongst compact objects, the typical amplitude of variations is often small compared to the factor of several variations that may be expected of the transients themselves. At centimetre wavelengths the expected refractive modulation index (i.e. root-mean-square flux density variation normalised by the mean flux density) is typically $\sim 0.1$, with an associated timescale of order days to weeks (e.g.\,Walker 1998).  For instance, the refractive scintillations exhibited by compact intraday-variable quasars typically display modulation indices less than $0.05$ (Lovell et al.\,2008) and timescales $>3$\,days, with timescales $>1-2\,$weeks at wavelengths longer than $\sim 10\,$cm (Lovell et al.\,2008; but see Kedziora-Chudczer et al.\,1997 \& Macquart \& de Bruyn 2006).  The timescale of refractive scintillation increases as $\lambda^2$, and its amplitude decreases with wavelength.  The effects of refractive scintillation on the detectability of transients are further discussed in \S\ref{RefISSeffect}.
   

\subsection{Event decoherence time} \label{sec:decoherence}


A consideration in a survey for slow transients is the desire to avoid repeatedly redetecting the same long-duration events.  We introduce the concept of the event decoherence function, which characterises how the particular transient events that are detected in a given survey field are correlated in time with those detected in a subsequent observation\footnote{Obviously, if all events are of identical duration the decoherence time is a trivial concept.  However, the function acquires practical significance when there is a distribution of event durations.}.  This quantity then determines the timescale over which successive visits to a given patch of sky would detect a fundamentally new set of transient sources.  Where redetection is deemed desirable, the event decoherence function, if known, allows one to set the reobservation time according to the fraction of events to be redetected.  

We represent the occurrence of an event by a function $f(t;\Delta T)$ such that, when the event occurs $f$ assumes the value one, and it is zero otherwise.  For an event that commences at $t=0$ and has duration $\Delta T$ one has $f(t) = H(t) - H(t-\Delta T)$, where $H(t)$ is the Heaviside function.  One may use this function to count the number of events occurring over a time interval ${\cal T}$ from the superset containing all ${\cal N}$ events.  For events due to a large ensemble of sources with event times $t_i$ and durations $\Delta T_i$, the number of transient events occurring at a time $t$ is given by
\begin{eqnarray}
N(t) = \sum_i^{\cal N} f(t-t_i; \Delta T_i). \label{Ndefn}
\end{eqnarray}
As we are interested in the correlations between different events, the quantity of interest here is the autocorrelation in the fluctuation in the event count, $\delta N(t) = N(t) - \bar{N}$.  For events which are distributed randomly and independently in time, this is given by (see Appendix \ref{ACFApp}),
\begin{eqnarray}
\langle \delta N(t+t') \delta N(t) \rangle &=& \left[ \frac{1}{2 {\cal T}} \sum_j^{\cal N} \left( |t+\Delta T_j |  + |t-\Delta T_j  | - 2 |t| \right) \right]  - {\cal O} \left( \frac{{\cal N} \langle \Delta T\rangle^2}{{\cal T}^2} \right),
\label{Nacf0}
\end{eqnarray}  
where we henceforth neglect the terms that are second order in ${\cal T}^{-1}$ or higher. We may satisfy ourselves that the formalism introduced here, in terms of event counting, is truly a measure of the timescale over which the detection of a set of events in a field becomes statistically independent from a future measurement.  An alternative conceptual approach is to consider the correlation in time between each event and every other event, and then average over all correlations of event pairs.  This is equivalent to an average over all pairs of events, and it is evident (see eq.\,(\ref{BigAvg})) that this is mathematically identical to the approach adopted here.

In order to evaluate the decoherence function explicitly we specify a probability distribution for the event durations, $\Delta T_i$ and take the continuum limit so that the sum over events becomes an integral.  Let us denote $p(\Delta T) d \Delta T$ as the probability of obtaining events between durations $\Delta T$ and $\Delta T + d \Delta T$.  Then the average event decoherence time is represented in the form,
\begin{eqnarray}
\langle \delta N(t+t') \delta N(t) \rangle = \left[ \frac{1}{\cal T} \int p(\Delta T)  \left[ (t'+\Delta T)  H(-t') H(t'+\Delta T) + (\Delta T - t') H(t')  H(\Delta T - t') \right]  d\Delta T \right]. \label{Nacf}
\end{eqnarray}
To acquire an intuitive appreciation of the event decoherence timescale, we evaluate the event decoherence function for three representative cases: events all possessing identical durations, events whose durations follow a gaussian distribution, and event durations that follow a power law distribution.

\subsubsection{Single event duration}
For a single event duration one has $p(\Delta T) = \delta (\Delta T - \Delta T_0)$, where $\Delta T_0$ is the duration of all the events.  In this case eq.\,(\ref{Nacf}) becomes 
\begin{eqnarray}
\langle \delta N(t+t') \delta N(t)  \rangle = \frac{{\cal N}}{2 {\cal T}}  \left( |t+\Delta T_0 |  + |t-\Delta T_0  | - 2 |t| \right) .
\end{eqnarray}
This illustrative result shows that, as intuitively expected, the correlation between events is a maximum at $t' = 0$ and falls to zero on a timescale of precisely $\Delta T_0$, the event duration.  The amplitude of the autocorrelation is linearly proportional to the event rate, ${\cal N}/{\cal T}$.  

\subsubsection{Gaussian duration distribution}
For a gaussian distribution of event durations with mean $\Delta T_0$ and standard deviation $\sigma_{\Delta T}$,
\begin{eqnarray}
p(\Delta T)  = \frac{1}{\sqrt{2 \pi \sigma_{\Delta T}^2}} \exp \left[ -\frac{(\Delta T-\Delta T_0)^2}{2 \sigma_{\Delta T}^2} \right],
\end{eqnarray}
one has 
\begin{eqnarray}
\langle \delta N(t+t') \delta N(t) \rangle =  \frac{{\cal N}}{{\cal T}} \begin{cases}
 G(\Delta T_0 - t') (\Delta T_0 - t')   + \sigma_{\Delta T}^2 g(\Delta T_0 - t'),  & t' > 0, \\
 G(\Delta T_0 + t') (\Delta T_0 + t')   +\sigma_{\Delta T}^2 g(\Delta T_0 + t'), & t' < 0, \\
\end{cases} \label{gaussACF}
\end{eqnarray}
where
\begin{eqnarray}
G(t) = \frac{1}{2} \left[1 + {\rm erf} \left( \frac{t}{\sigma_{\Delta T} \sqrt{2}} \right) \right]  \quad \hbox{and} \quad g(t) = \frac{1}{\sqrt{2 \pi \sigma_{\Delta T}^2}} \exp \left[- \frac{t^2}{2 \sigma_{\Delta T}^2} \right]
\end{eqnarray}
are the cumulative and probability density functions of a gaussian distribution respectively.

The width of this autocorrelation is determined by both $\Delta T_0$ and $\sigma_{\Delta T}$.  When $\Delta T_0$ is large compared to $\sigma_{\Delta T}$, the second term in eq.(\ref{gaussACF}) is small, and the behaviour of the autocorrelation is dominated by the first term.  For $\sigma_{\Delta T}$ small, the function in the first term, $1-{\rm erf}[(t'-\Delta T_0)/(\sigma_{\rm \Delta T} \sqrt{2})]$, behaves like $2 H(t'-\Delta T_0)$.  The result is an autocorrelation function which is dominated by a linear decrease with $t'$ until the autocorrelation reaches zero at $t'=\Delta T_0$.

As $\sigma_{T}$ becomes larger one finds that the second term in eq.(\ref{gaussACF}) increases in magnitude, and its effect is to add gaussian wings to the autocorrelation at $t'>\Delta T_0$. We note, however that, for the purposes of this particular distribution, it is unphysical to consider situations in which $\sigma_{T} > \Delta T_0$ because the probability distribution of event durations would then imply that a large number of events with negative durations occur.  

The event decoherence function declines linearly for lags  $t' \la \Delta T_0$ and then declines exponentially quickly to zero for lags $t' \ga \Delta T_0$.

\subsubsection{Power-law event duration distribution}
Another likely possibility is one in which the distribution of event durations follows a power law of the form,
\begin{eqnarray}
p(\Delta T) = k \Delta T^{-\gamma},  \quad k=\frac{1-\gamma}{\Delta T_{\rm max}^{1-\gamma} - \Delta T_{\rm min}^{1-\gamma}} \hbox{ for } \gamma \neq 1,
\end{eqnarray}
for durations between $\Delta T_{\rm min}$ and $\Delta T_{\rm max}$.   For such an event duration distribution, the event decoherence function takes the form,
\begin{eqnarray}
\langle \delta N(t+t') \delta N(t) \rangle = \frac{{\cal N}}{{\cal T}} \begin{cases}
F(\Delta T_{\rm min}) - t', 
	& 0< t' < \Delta T_{\rm min}, \\
F(t') - t' f(t'), 
	& \Delta T_{\rm min} < t' < \Delta T_{\rm max}, \\
0, 
	& t' > \Delta T_{\rm max}, \\
\end{cases}
\end{eqnarray}
where we define
\begin{eqnarray}
F(t) = \frac{\gamma-1}{\gamma-2}\left[ \frac{\Delta T_{\rm max}^{2-\gamma} - {t}^{2-\gamma} }{\Delta T_{\rm max}^{1-\gamma} - \Delta T_{\rm min}^{1-\gamma}} \right]  
\quad \hbox{and} \quad f(t) = \left[ \frac{\Delta T_{\rm max}^{1-\gamma}-{t}^{1-\gamma}}{\Delta T_{\rm max}^{1-\gamma}-\Delta T_{\rm min}^{1-\gamma}} \right].
\end{eqnarray}
It is obvious that this autocorrelation falls to zero on a timescale $t' = \Delta T_{\rm max}$.  However, as the value of $\gamma $ increases, more of the event durations are close to $\Delta T_{\rm min}$, and the decoherence function is more sharply peaked about the origin; accordingly, the half power point of the  function moves to progressively smaller time lags and approaches $\Delta T_{\rm min}$.

\begin{figure*}[h!]
\centerline{\epsfig{file=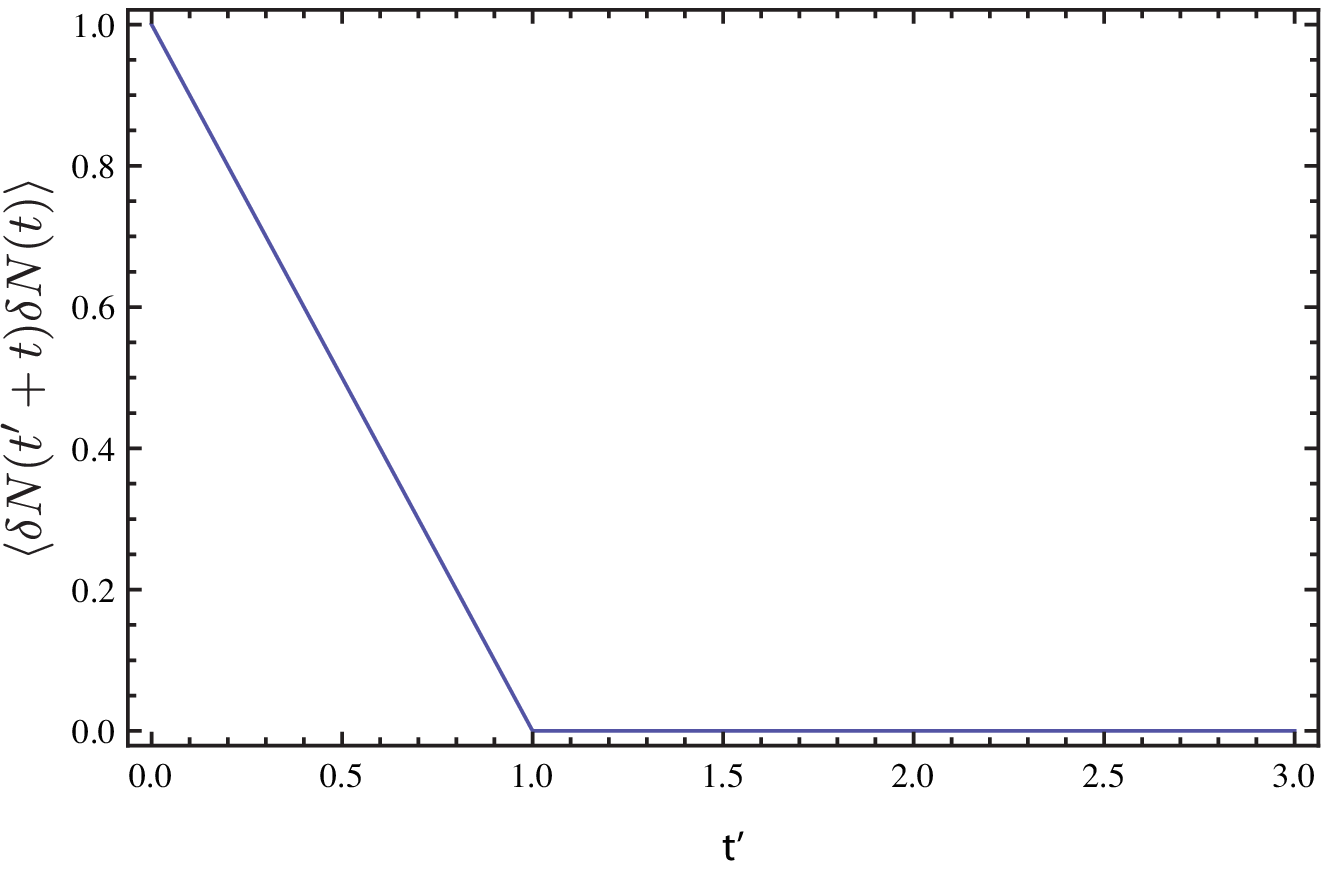,scale=0.38}
\epsfig{file=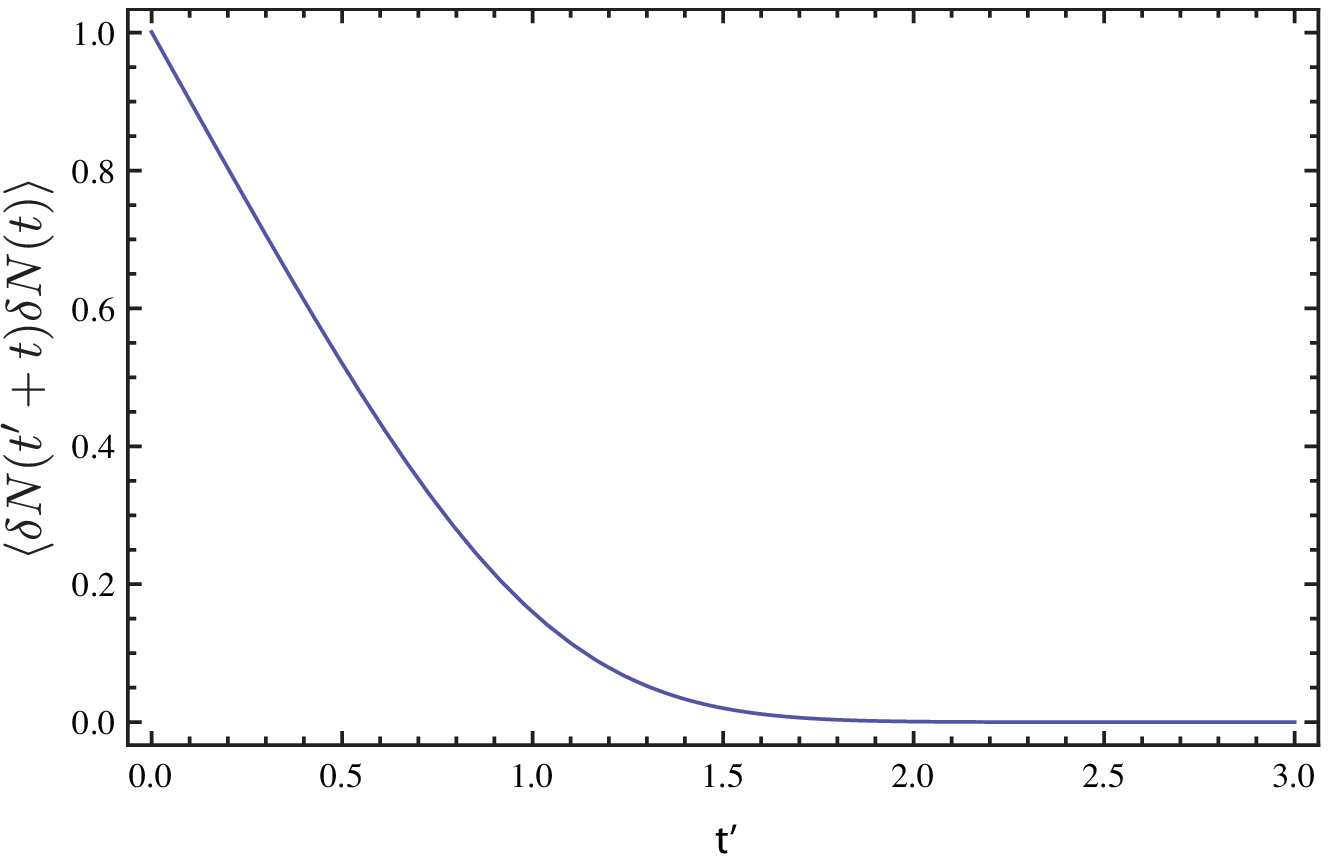,scale=0.38}
\epsfig{file=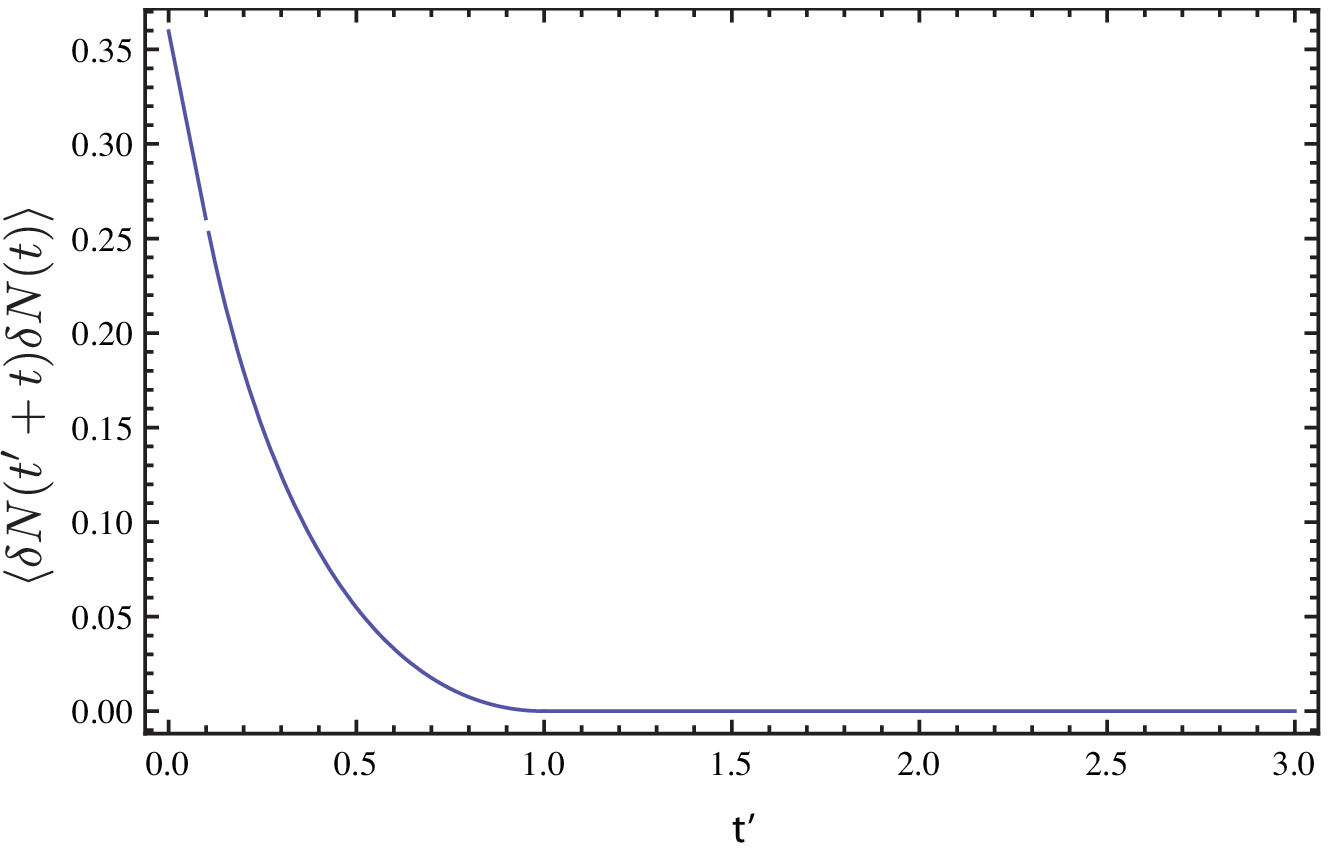,scale=0.38}}
\caption{The decoherence function for three possible event duration distributions, normalised by the event rate $N/{\cal T}$.  The plots show the decoherence function, from left to right, for (i) a set of events all with the same duration $\Delta T = 1$, (ii) a Gaussian duration distribution with $\Delta T_0=1$ and $\sigma_{\Delta T} = 0.4$, and (iii) a power law duration distribution with $\Delta T_{\rm min} = 0.1$, $\Delta T_{\rm max}=1$ and $\gamma=1.2$.} \label{figACFs}
\end{figure*}

Illustrative plots of the event decoherence function for the various duration probability distribution functions examined here are shown in Figure \ref{figACFs}.  Hereafter throughout the text we denote $T_{\rm decoher}$ as the timescale at which the event decoherence function effectively decays to a negligible value.  

\subsection{Event types} 

An important consideration in the optimization of a transients survey relates to the types of events one should optimize for.  Our analysis is informed by two premises:
\begin{itemize}
\item[i] The average event rate is uniform across the survey region on the sky and is homogeneous in time. 
\item[ii] The events that we are interested in optimizing the survey for occur on a duration longer than the duration that the telescope dwells on any one field.  This tenet relates to our definition of a slow transient; if the event duration is shorter than the telescope dwell time, no advantage is gained in event rate by slewing the telescope to survey another location.  Based on tenet (i), the event detection rate for these faster-timescale transients is identical whether one continues surveying the same patch of sky or moves to another field.  Thus the only events worth considering in an optimization of dwell time versus survey area in a slow transients are those whose duration exceeds the dwell time.  
\end{itemize}

Before determining how to optimize the event detection rate within a survey, it is first necessary to ascertain the rate at which a telescope with a given field of view, $\Omega$, and sensitivity would detect transients in a single integration on a single field.   We consider a survey with a telescope whose limiting detection flux density within an observing duration $T_{\rm obs}$ is, 
\begin{eqnarray}
S_0 = \frac{m k_B T_{\rm sys}}{A_e \sqrt{n_p \Delta \nu T_{\rm obs}}},
\end{eqnarray}
where $\Delta \nu$ is the observing bandwidth, $A_e$ the telescope effective area, $T_{\rm sys}$ the system temperature, $n_p$ the number of polarizations recorded and $m$ the minimum signal-to-noise ratio that is acceptable for detection (e.g. $m=7$ for a 7$\sigma$ detection)\footnote{We assume throughout this work that the sensitivity of the observation is dominated by thermal receiver noise, as is appropriate for a many-element interferometer with excellent $uv$ coverage, such as MeerKAT and the SKA.  Imaging artefacts may instead limit the practical sensitivity achievable, especially if the $uv$ coverage of the radio array is sufficiently poor over a time $\sim T_{\rm obs}$.  In this case one might expect that the limiting ``believable'' flux density, $S_0$ will decrease more slowly than the $T_{\rm obs}^{-1/2}$ dependence of thermal noise.  The rate at which this occurs will depend heavily on the array configuration and sky complexity at the observing frequency, and a general analysis of its effects for specific arrays is beyond the scope of this paper.  However, the analysis presented here is readily generalised to take into account such effects if their average dependence on integration time is known.}.   

Some discussion regarding point (ii) above is in order.  In practice, the survey will detect events of durations both shorter and longer than the telescope dwell time.
For those events with durations whose spans smaller than $T_{\rm obs}$, the sensitivity to an event of duration $t_{\rm tr}$ is smaller than $S_0$ by a factor $\sqrt{T_{\rm obs}/t_{\rm tr}}$, and the detection rate of these ``fast'' transients will depend in detail on the bi-variate distribution of event luminosity and duration relative to the telescope limiting flux density (which will vary for each event depending on its duration).  Thus the survey detection rate for such events is a complicated function of the particular event luminosity-duration distribution that pertains to any given transient population, which is hard to predict {\it a priori}.  Fortunately, however, by tenet (ii) there is no advantage to gained by trading integration time versus survey area for such transient events.  An important point is that the survey will detect just as many of these fast transients by observing the same field as it would by observing another field.  As such events make no contribution to the survey rate optimization, we do not consider the contributions of events $t_{\rm tr} < T_{\rm obs}$, further in the present analysis, and our detection rate does not take them into account.  The reader is referred to Macquart (2011) for an analysis of their contribution to the event rate.

In this work we consider events whose luminosity function follows a power law with index $-\alpha$ between luminosities $L_{\rm min}$ and $L_{\rm max}$ such that the volume density of objects between luminosities $L$ and $L+dL$ is
\begin{eqnarray}
\rho_L dL = \frac{\rho_0 dL}{K} \left\{ 
\begin{array}{ll} 
L^{-\alpha}, & L_{\rm min} < L < L_{\rm max}, \\
0, & \hbox{otherwise},
\end{array}
\right. \label{Ldefn}
\end{eqnarray}
where $\rho_0$ is the total event rate per unit volume integrated over all luminosities and the normalization constant is
\begin{eqnarray}
K = \left\{ 
\begin{array}{ll}
\frac{1}{1- \alpha  } \left( L_{\rm max}^{1-\alpha} - L_{\rm min}^{1-\alpha} \right) , & \alpha > 0 \hbox{ and } \alpha \neq 1, \\
\ln \left( \frac{L_{\rm max}}{L_{\rm min}} \right) & \alpha =1. \\
\end{array}
\right.
\end{eqnarray}

\section{Detection Rate} \label{SingleField}

We consider the event detection rate for two types of surveys.  The first relates to a search for events whose progenitors are distributed homogeneously throughout space, and the second relates to a targetted search of a system (e.g. a galaxy or a cluster of galaxies) located at a fixed distance from the observer.  We additionally consider an obvious generalization of the former case in which the events are distributed homogeneously throughout space, but are only detectable beyond a certain minimum distance; this case illustrates the effect on survey rate of a population whose detection requires a certain minimum sensitivity. 

\subsection{Homogeneously distributed events}
For a population of transients distributed homogeneously throughout space, the telescope is capable of detecting objects of luminosity $L$ out to a luminosity distance $D_{\rm max} = \sqrt{ L/4 \pi S_0}$, so that the observed number of events per second for objects of of luminosities between $L$ and $L+dL$ is, 
\begin{eqnarray}
{\cal R} dL = \rho_L dL \frac{\Omega}{4 \pi} \frac{4}{3} \pi D_{\rm max}^3,
\end{eqnarray}
where $\Omega$ is the telescope field of view.  The total event rate integrated over all luminosities may be expressed in the compact form\footnote{This result is valid for a Euclidean space geometry only, but can be generalized if the events under consideration occur at cosmological distances (i.e. $z \gtrsim 1$).  This generalization is discussed in Appendix \ref{AppendixCosmo}.  However, in many populations at cosmological distances (e.g. quasars), strong evolution in the source population overwhelms cosmological effects when considering the event rate.  We do not attempt to model the cosmological evolution of any hypothetical population of transient progenitors in this work.},
\begin{eqnarray}
&\null& R_{\rm total} = A \, \Omega \, T_{\rm obs}^{3/4}, \label{RateHomog} \\
&\null& \hbox{ where }  A =  \frac{\rho_0 }{3 \sqrt{4 \pi} \beta_0^{3/2}} \frac{1}{4 \pi} 
\left( \frac{1-\alpha}{\frac{5}{2}-\alpha} \right)
\frac{L_{\rm max}^{\frac{5}{2}-\alpha}-L_{\rm min}^{\frac{5}{2}-\alpha}}{L_{\rm max}^{1-\alpha}-L_{\rm min}^{1-\alpha}}, \,\,  \alpha \neq \left\{1, \frac{5}{2} \right\} \hbox{ and } S_0 = \frac{\beta_0}{\sqrt{T_{\rm obs}}}. \nonumber \\ 
\end{eqnarray}
It is apparent that the detection rate scales linearly with field of view, but increases more slowly with telescope dwell time.  This is the basis of the survey optimization: the event detection rate may be increased by sacrificing telescope dwell time in favour of field of view by make multiple pointings over a given survey time $T_{\rm obs}$.  

\subsubsection{Homogeneously distributed events only detectable beyond a minimum distance}
The event rate in a survey over a truly homogeneously-distributed population is always nonzero, however how poor the survey sensitivity, by virtue of the fact that the survey may detect events arbitrarily close to the observer.  

In certain cases it is more realistic to suppose events are only detectable beyond a certain minimum distance, $D_{\rm min}$.  This situation effectively pertains to searches in which a certain minimum sensitivity is required before the survey can detect far enough out into the Universe in order to detect any of the target events.  

The detection rate between luminosities $L$ and $L + dL$ is therefore
\begin{eqnarray}
R dL = \rho_L dL \frac{\Omega}{4 \pi} \left(\frac{4 \pi}{3} D_{\rm max}^3 - \frac{4 \pi}{3} D_{\rm min}^3 \right), \quad D_{\rm max} > D_{\rm min}.
\end{eqnarray}
The requirement $D_{\rm max} > D_{\rm min}$ implies an inequality between luminosity, sensitivity and $D_{\rm min}$: $\sqrt{L/4 \pi S_0} > D_{\rm min}$, and the detection rate integrated over all luminosities is
\begin{eqnarray}
R = \frac{\Omega}{4 \pi} \frac{4 \pi}{3} \frac{\rho_0}{K} \int_{L_{\rm min}}^{L_{\rm max}} L^{- \alpha} 
\left[ \left( \frac{L}{4 \pi S_0} \right)^{3/2} - D_{\rm min}^3 \right] H\left[ L - 4 \pi S_0 D_{\rm min}^2 \right] dL. 
\end{eqnarray}
The behaviour of the detection rate depends on the survey sensitivity relative to the flux density of the brightest and faintest objects detectable at the inner radius of the survey volume.  
The detection rate is, 
\begin{eqnarray}
R = \frac{\Omega \rho_0}{3} 
\begin{cases}
\frac{\zeta(L_{\rm min})}{(4 \pi S_0)^{3/2}}  - D_{\rm min}^3 
	, & \frac{L_{\rm min}}{4 \pi D_{\rm min}^2} >  S_0,  \\
\frac{\zeta(L')}{(4 \pi S_0)^{3/2}}  
-  \eta(L')  D_{\rm min}^3 
	, & \frac{L_{\rm min}}{4 \pi D_{\rm min}^2} < S_0  < \frac{L_{\rm max}}{4 \pi D_{\rm min}^2}, \\
0, 
	&  \frac{L_{\rm max}}{4 \pi D_{\rm min}^2} < S_0. \\ 
\end{cases} \label{RminDist}
\end{eqnarray}
where it is convenient to define,
\begin{eqnarray}
\zeta(L) =  \left(\frac{1-\alpha}{\frac{5}{2}-\alpha}\right) \left( 
\frac{L_{\rm max}^{5/2-\alpha} - {L}^{5/2-\alpha} }{L_{\rm max}^{1-\alpha} - L_{\rm min}^{1-\alpha}}  \right) \quad \hbox{and} \qquad 
\eta(L) = \left( \frac{L_{\rm max}^{1-\alpha} - {L}^{1-\alpha} }{L_{\rm max}^{1-\alpha} - L_{\rm min}^{1-\alpha}}  \right), 
\end{eqnarray}
and where we define the lowest luminosity detectable at the inner radius of the survey volume as $L^{\prime} = 4 \pi S_0 D_{\rm min}^2$.  These results may be understood as follows.  If the survey sensitivity is so poor that $S_0 > L_{\rm max}/4 \pi D_{\rm min}^2$, it is incapable of even detecting events at the minimum distance, and the event rate is zero.  If the survey is extremely sensitive, $ S_0 < L_{\rm min}/4 \pi D_{\rm min}^2$,  the survey detects every event at the inner radius of the survey volume, and the event rate is the same as that for a survey of homogeneously-distributed events except that there is no contribution from events interior to the volume bounded by the radius $D_{\rm min}$.  In the intermediate case, $L_{\rm min} < L^{\prime} < L_{\rm max}$, the survey only detects that fraction of the events whose luminosities exceed $L^\prime$ at the inner radius of the survey volume.

\subsection{A targeted survey of a system at fixed distance}
Another possible survey mode involves targeted observations of a galaxy at a known luminosity distance, $D_L$.  Since all the putative transient events are located at the same distance, we detect all events whose luminosity exceeds $L_0 = 4 \pi S_0 D_L^2$.
If the total event rate per unit volume is $\rho_0$, then the total event rate per unit solid angle, $\rho_\Omega$ is found by integrating over the depth of the object, $\Delta z$ (assumed small compared to its distance) and over the surface area subtended by the portion of sphere of radius $D_{\rm comoving}$ that encompasses the telescope field of view.  An observation over a solid angle $\Omega$ would detect a fraction $\Omega/4 \pi$ of these objects over the total surface $4 \pi D_{\rm comoving}^2$ and subtend an area $\Omega D_A^2$, where $D_A=D_{\rm comoving}/(1+z)$ is the angular diameter distance to the object and $D_L = (1+z)^2 D_A$.  Thus the total event rate per solid angle is 
\begin{eqnarray}
\rho_\Omega = \rho_0 \Delta z D_A^2 \Omega,
\end{eqnarray}
and so we identify the rate of events between luminosities $L$ and $L+dL$ as 
\begin{eqnarray}
\rho_{L,\Omega} dL = \frac{\rho_0 \Delta z D_A^2 \Omega }{K}  L^{-\alpha} dL =  \frac{\rho_0 \Delta z D_A^2 \Omega (1-\alpha) }{L_{\rm max}^{1-\alpha} - L_{\rm min}^{1-\alpha}} L^{-\alpha} dL. \label{Ldist}
\end{eqnarray}

\begin{figure}
\centerline{\epsfig{file=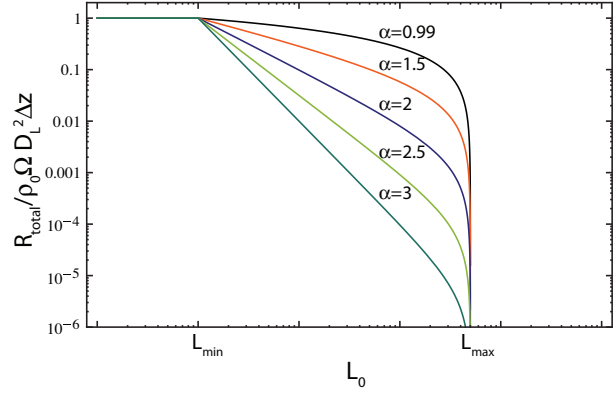, scale=0.6}}
\caption{A schematic illustration of the behaviour of the event detection rate, $R_{\rm tot}$ as a function of the limiting luminosity, $L_0$, to which events can be detected in a survey of a system at a fixed distance.  If $L_0$ is sufficiently low, we detect all objects in the galaxy, but if $L_0 > L_{\rm max}$ we detect no events.  This plot shows the event rate for a luminosity function for various values of $\alpha$.} \label{RtotDemo}
\end{figure}

The total observed detection rate is therefore
\begin{eqnarray}
R_{\rm total} &=&  \int_{{\rm max}[L_0,L_{\rm min}]}^{{\rm min}[L_0,L_{\rm max}]} \rho_{L,\Omega} dL  
= \rho_0 \Delta z D_A^2 \Omega \left\{ 
\begin{array}{ll}
1, & L_0 < L_{\rm min}, \\
\eta(L_0), & L_{\rm min} < L_0 < L_{\rm max}, \\
0, & L_0 > L_{\rm max}. \\
\end{array}
\right. \label{RateTarget}
\end{eqnarray}
The three cases in the above expression are interpreted to mean: (i) when $L_0 < L_{\rm min}$ the sensitivity of the observation is so good that even the faintest event can be detected by our observation, so we detect all possible events and the measured event rate is equal to the intrinsic field event rate of $\rho_0 \Delta z D_A^2 \Omega$; (ii) when $L_0 > L_{\rm max}$, the sensitivity of the observation is so poor that not even the most luminous event can be detected, so the event detection rate is zero;  and (iii) for the case $L_{\rm min} < L < L_{\rm max}$, the sensitivity is intermediate to the two preceding cases and we detect only a fraction $\frac{L_{\rm max}^{1-\alpha} - L_0^{1-\alpha}}{L_{\rm max}^{1-\alpha} - L_{\rm min}^{1-\alpha}}$ of all the events that occur.  An illustration of the behaviour of $R_{\rm total}$ as a function of $L_0$ is shown in Figure \ref{RtotDemo}.

\subsubsection{Correction for the effect of refractive interstellar scintillation} \label{RefISSeffect}

As an illustration of the effects of interstellar scintillation on the results presented in this paper, we consider briefly here how interstellar scintillation 
alters the foregoing result.  
Interstellar scintillation can randomly amplify or deamplify the signals of compact transients, and so alter the apparent distribution of event flux densities.  In situations where the scintillation timescale is long compared to both the event duration and the duration of the observation, scintillation can bear on the detection rate by amplifying some events that would otherwise be too faint to detect.

\begin{figure*}[h!]
\centerline{\epsfig{file=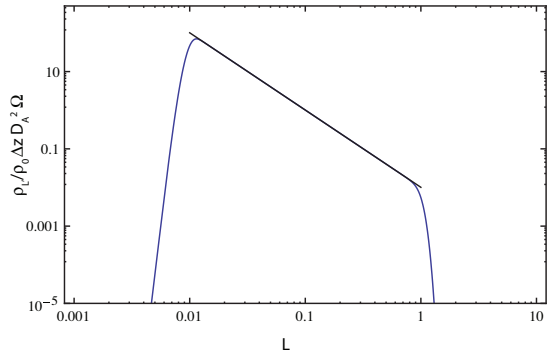,scale=0.55}}
\caption{The effective luminosity function (blue curve) of eq.\,(\ref{Leff}) for $m=0.1$, $L_{\rm min}=0.01$ and $L_{\rm max}=1$.  Overplotted in black is the intrinsic luminosity function, $\rho_L$.} \label{FigScint}
\end{figure*}

For events all located at the same distance, as considered in this subsection, there is a linear relationship between source luminosity and flux density.  It is then straightforward to incorporate this scintillation effect by considering an ``effective'' event luminosity distribution that replaces that given in eq.(\ref{Ldist}) by making the replacement $L' = a L$, where $a$ is the amplification caused by scintillation.  The resulting distribution of effective event luminosities is then given by,
\begin{eqnarray}
\rho_{L'} = \int_{-\infty}^\infty p_a(a) \rho_L \left( \frac{L}{a} \right)\frac{da}{|a|},
\end{eqnarray}
where $p_a$ is the probability distribution of amplifications, $\rho_L$ is the distribution of intrinsic luminosities (viz. eq.\,(\ref{Ldefn})), and $\rho_{L'}$ is the distribution of effective event luminosities after scintillation is taken into account.  For the moderate modulation indices typical of refractive scintillation at centimetre and decametre wavelengths (i.e.\,$m$ much less than one), the amplification probability distribution takes the form,
\begin{eqnarray}
p_a(a)  = \frac{1}{\sqrt{2 \pi m^2}} \exp \left[ - \frac{(a-1)^2}{2 m^2} \right].
\end{eqnarray}
We assume that the timescale of both the event and of the observation is short compared to the scintillation timescale; if the observed emission from the transient encompasses several (say $n$) independent scintillations the modulation index is reduced by a factor of $n^{1/2}$.  Noting that $\rho_L$ is non-zero only over the range $[L_{\rm min},L_{\rm max}]$, the effective luminosity function becomes,
\begin{eqnarray}
\rho_{L'}  = \frac{1}{\sqrt{2 \pi m^2}}  \int_{L/L_{\rm max}}^{L/L_{\rm min}} \frac{da}{a} \exp \left[ - \frac{(a-1)^2}{2 m^2} \right] \, p_L \left( \frac{L}{a} \right).
\end{eqnarray}
We are unable to evaluate this integral in closed form for non-integer values of $\alpha$, however we may still understand the effects of scintillation qualitatively by considering its solution for integer values.  The solution for $\alpha=2$ is particularly simple:
\begin{eqnarray}
\rho_{L'} &=& \frac{\rho_0 \Delta z D_A^2 \Omega L^{-2}}{L_{\rm min}^{-1} - L_{\rm max}^{-1}} 
\left\{ \frac{1}{2} 
	\left[ {\rm erf} \left( \frac{L-L_{\rm min}}{\sqrt{2} \, m L_{\rm min}} \right) - 
	{\rm erf} \left( \frac{ L-L_{\rm max} }{\sqrt{2} \, m  L_{\rm max}}\right) \right]  \right. \nonumber \\
&\null& \left.\qquad \qquad \qquad - \frac{m}{\sqrt{2 \pi}} 
	\left[ \exp \left( - \frac{( L - L_{\rm min})^2}{2 m^2 L_{\rm min}^2} \right) 
	- \exp \left( - \frac{ (L - L_{\rm max})^2}{2 m^2 L_{\rm max}^2 }  \right) \right]  \right\}. \label{Leff}
\end{eqnarray}
Scintillation only alters the shape of $\rho_{L'}$ near $L = L_{\rm min}$ and $L=L_{\rm max}$; for luminosities in the range $[L_{\rm min} (1+\sqrt{2} m),L_{\rm max} (1-\sqrt{2} m) ]$ the effective luminosity distribution is, to an excellent approximation, $\rho_{L'} = L^{-2}/(L_{\rm min}^{-1} - L_{\rm max}^{-1})$.  A plot of the behaviour of this function is shown in Figure\,\ref{FigScint}.  It shows that the effect of scintillation is to extend the low and high flux density limits of the distribution to yet lower and higher values respectively.  The behaviour of the effective luminosity distribution function near $L_{\rm min}$ and $L_{\rm max}$ is dominated by the error functions in eq.(\ref{Leff}), from which we deduce that the effect of scintillation is to extend a small fraction of high luminosity events to yet higher luminosities, 
$\approx L_{\rm max}( 1+ \sqrt{2}  m)$, and some events at the low luminosity end down to values $\approx L_{\rm min} (1- \sqrt{2} m)$.

In practice, for the small modulation indices ($m < 0.1$) typically observed for refractive interstellar scintillation, the effects of scintillation on the event detection rate are small unless the survey sensitivity is such that the limiting luminosity of the observations, $L_0, $ is close to $L_{\rm max}$.  We note, however, that there are cases in which the refractive modulation index is observed to be higher ($\approx 0.3$), and that compact objects are also subject to the effects of  Extreme Scattering Events (ESEs; Fiedler et al.\,1987).   ESEs may produce flux density deviations exceeding 50\% in some cases, but these events are rare, estimated at 0.013 src$^{-1}$\,year$^{-1}$.

\section{Optimising the Detection Rate for Surveys over Multiple Fields} \label{MultiFields}
In the section above we have derived the detection rate for a single telescope field of solid angle $\Omega$ whose sensitivity is such that it can detect objects down to a limiting flux density of $S_0$ in a duration $T_{\rm obs}$.  In this section we consider the detection rate for a modified search strategy in which the total observing time is subdivided so that we visit a number of fields, $N$.  If the telescope requires a time  $t_{\rm slew}$ to move between fields, then the total amount of time spent slewing is\footnote{This term is proportional to $N-1$ since a slew to the first field is unnecessary and is not included in the single-field event rate calculations of eqs.(\ref{RateHomog}) and (\ref{RateTarget}): the exact starting position is unimportant in a survey for events distributed homogeneously (and thus isotropically about the sky).   However, where time consumed in an initial slew is an important consideration, one may always choose to reinterpret $N$ as $N= N_{\rm extra\,slew} +1$.} $(N-1) t_{\rm slew}$, and the total time spent observing on each field is
\begin{eqnarray}
T_{\rm field} = \frac{T_{\rm obs} - (N-1) t_{\rm slew}}{N}.
\end{eqnarray}
The following results are also applicable to drift-scan surveys, in which case $t_{\rm slew}$ should be identified as the time that it takes the telescope beam to traverse a fixed point in the sky.

Although the survey covers more field of view by increasing $N$, the limiting sensitivity per field will be commensurately lower, and it is our purpose to determine the value of $N$ that maximizes the total number of detections over the interval $T_{\rm obs}$.

\subsection{Homogeneously distributed objects}
Equation (\ref{RateHomog}) presents the event detection rate for a telescope whose sensitivity is such that it can detect objects down to a limiting flux density of $S_0$ in a duration $T_{\rm obs}$.   The event detection rate for a survey over $N$ fields for the same duration is then
\begin{eqnarray}
R_{\rm N-fields} = A \, N \Omega \, \left( 
\frac{T_{\rm obs} - (N-1) t_{\rm slew}}{N} \right)^{3/4},
\end{eqnarray}
and the maximum event rate occurs when the number of fields visited is
\begin{eqnarray}
N_{\rm max} = \frac{T_{\rm obs}+ t_{\rm slew}}{4 t_{\rm slew}}. \label{Nmax}
\end{eqnarray}
The associated maximum event rate is 
\begin{eqnarray}
R_{\rm max} = \frac{3^{3/4} \,A \, \Omega }{4} t_{\rm slew}^{-1/4} \left(T_{\rm obs} + t_{\rm slew} \right),
\end{eqnarray}
which is a factor $\sim 0.57 (T_{\rm obs}/t_{\rm slew})^{1/4}$ larger than the event rate that would be obtained by spending the entire integration time on a single field.

A few obvious points are in order.  We have regarded $N$ as a continuous variable, but in practice $N$ is an integer, so it will only be possible to approximately obtain the maximum event rate.  Where $N_{\rm max}$ is less than one, the optimal strategy is to survey only a single field.

Equation (\ref{Nmax}) indicates that, as $T_{\rm obs}$ increases, the number of fields that should be searched increases similarly.  However, there is a practical limitation to the number of fields to be searched which is set by the physics of the target transient events.  After a duration comparable to the event decoherence time one expects each of the searched fields to contain a new set of transient events, so that a second measurement of the same field would be statistically independent of the original measurement.  Thus a search of the same field after a duration $T_{\rm decoher}$ is statistically equivalent to searching another (as yet) unsearched field.  As there is no advantage to searching other statistically independent fields over the same fields a second time, this allows us to identify a practical search cycle time as $T_{\rm decoher}$.  Thus the maximum number of fields that the survey should cycle over is $N_{\rm max} = (T_{\rm decoher} + t_{\rm slew})/4 t_{\rm slew}$.

\subsubsection{Homogeneously distributed objects detectable only beyond a minimum distance}
\begin{figure*}[h!]
\begin{center}
\epsfig{file=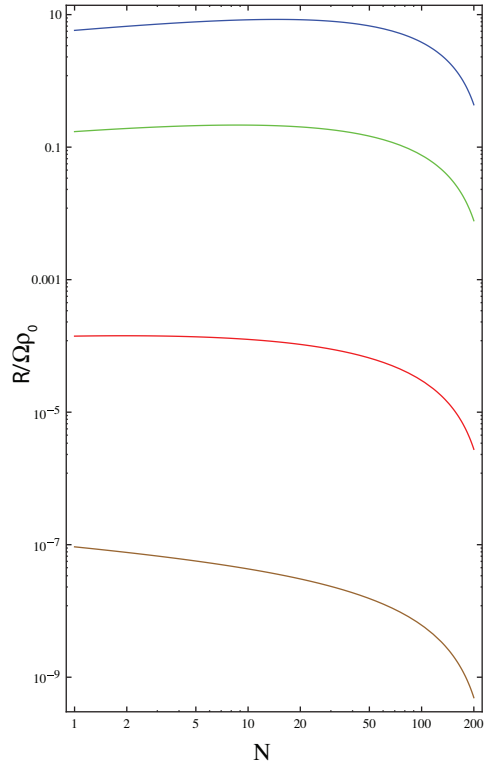, scale=0.5}
\end{center}
\caption{A plot of the event detection rate for the case in which the events are distributed homogeneously throughout space, but are only detectable beyond some minimum distance from the observer, $D_{\rm min}$.  Here we have chosen $L_{\rm min} =0.001$, $L_{\rm max}=10$, $S_0 = 0.05$ and  $D_{\rm min}=1$.  Shown in different colours and on different scales are the event rate plots for $\alpha=0.4$ (blue), $\alpha=1.4$ (green), $\alpha=2.4$ (red) and $\alpha=3.4$ (brown).  Note that only the curves with $\alpha< 3$ exhibit a peak as function of $N$.  For larger values of $\alpha$ the detection rate is a monotonically decreasing function of $N$.} \label{figNfieldDmin}
\end{figure*}

Here we examine the optimum detection strategy for the slightly more complicated case in which the transient events are only detectable beyond a certain minimum distance.  For the sake of algebraic simplicity we assume that the slewing time is negligible relative to observing time, so that the total amount of time spent per field is $T_{\rm obs}/N$. In this case the detection rate is obtained from eq.(\ref{RminDist}) by making the replacement $\Omega \rightarrow \Omega N$ and $S_0 \rightarrow S_0 \sqrt{N}$:  
\begin{eqnarray}
{\footnotesize
R_{\rm N-fields} =  \frac{N \Omega \rho_0}{3} 
\begin{cases}
\frac{ \zeta(L_{\rm min}) }{(4 \pi N^{1/2} S_0)^{3/2}}  - D_{\rm min}^3 
	, & \frac{L_{\rm min}}{4 \pi D_{\rm min}^2} >  N^{1/2} S_0,  \\
\frac{ \zeta(N^{1/2} L') }{(4 \pi N^{1/2} S_0)^{3/2}}   
-  \eta (N^{1/2} L') D_{\rm min}^3 
	, & \frac{L_{\rm min}}{4 \pi D_{\rm min}^2} < N^{1/2} S_0  < \frac{L_{\rm max}}{4 \pi D_{\rm min}^2}, \\
0, 
	&  \frac{L_{\rm max}}{4 \pi D_{\rm min}^2} < N^{1/2} S_0 . \\ 
\end{cases} 
} \nonumber \\  \label{RDmin}
\end{eqnarray}
The behaviour of this function is shown in Figure \ref{figNfieldDmin}, from which it is clear that the maximum event rate occurs at $N>1$ for luminosity functions shallower than $\alpha =3$.  

We derive an exact expression for the value of $N$ that optimises the event rate by differentiating eq.\,(\ref{RDmin}). For most cases of interest, in which $L_{\rm max} \gg L_{\rm min}$, the maximum rate occurs in the range $L_{\rm min}/4 \pi S_0 D_{\rm min}^2  < N^{1/2} < L_{\rm max}/4 \pi S_0 D_{\rm min}^2$. The location of the maximum is given by the lowest nonzero solution to the following equation:
\begin{eqnarray}
0= 3 (\alpha-3) {L^{\prime}}^{5/2} N^{5/4} + (\alpha-1) {L^{\prime}}^{\alpha} L_{\rm max}^{5/2} N^{\alpha/2} - 2 (2 \alpha -5) {L^{\prime}}^{3/2+\alpha} L_{\rm max} N^{3/4+\alpha/2} . \label{N4solns} 
\end{eqnarray}
As is obvious from Figure \ref{figNfieldDmin}, the optimal number of survey fields decreases as the luminosity function becomes progressively steeper.   Equation (\ref{N4solns}) also has the solution $N = (L_{\rm max}/4 \pi S_0)^2$, which corresponds to the point at which the event detection rate reaches zero; for $\alpha >3$ this is the only nonzero solution, reflecting the fact that when $\alpha > 3$ the detection rate becomes a monotonically decreasing function of $N$, and the optimal strategy is to survey only a single field.

The turning point may, in principle, instead occur in the regime $N^{1/2} < L_{\rm min}/4 \pi S_0 D_{\rm min}^2$, in which case it occurs at a value of $N$ given by
\begin{eqnarray}
N = \left[ \frac{1}{2 D_{\rm min}^3 (4 \pi S_0)^{3/2}} \left( \frac{1-\alpha}{5/2-\alpha} \right) 
\frac{L_{\rm max}^{5/2-\alpha} - L_{\rm min}^{5/2-\alpha}}{L_{\rm max}^{1-\alpha} - L_{\rm min}^{1-\alpha}} \right]^2.
\end{eqnarray}
Combining this solution with the inequality, $N < (L_{\rm min}/L^{\prime})^2$, we see that this solution is only valid when
\begin{eqnarray}
 \zeta(L_{\rm min}) < 2 L_{\rm min} {L^{\prime}}^{1/2}.
\end{eqnarray}
In most practical situations this alternative solution may be ignored.

\subsection{A targeted survey of systems at fixed distance}

Let us now consider the detection rate over a set of targeted objects at fixed distance.  This situation is applicable to a survey of an individual galaxy or cluster of galaxies (e.g. the Virgo supercluster) whose angular size subtends more than one telescope beam, so that multiple pointings are required to cover it.  More generally, the situation is applicable to any survey of multiple targets which are approximately equidistant but located in separate fields of view.

We start from equation (\ref{RateTarget}), which relates the detection rate in a single field to the luminosity distribution of the transient population and our telescope sensitivity, expressed in terms of the limiting detectable luminosity, $L_0 = 4 \pi D_L^2 S_0$.  Now, since $S_0 \propto T_{\rm obs}^{-1/2}$, we see that the limiting luminosity of a survey of total duration $T_{\rm obs}$ which visits $N$ fields is approximately $\sqrt{N} L_0$.  If we account for slewing time, we see that the total time per observation is $(T_{\rm obs} - (N-1) t_{\rm slew})/N$, so the limiting luminosity increases to $L_{\cal N} = L_0 \sqrt{N T_{\rm obs}/[T_{\rm obs} - (N-1) t_{\rm slew}]}$.  The detection rate for a survey which visits $N$ fields is then,
\begin{eqnarray}
R_{\rm N-fields} &=& N \rho_0 \Delta z D_A^2  \Omega \left\{ 
\begin{array}{ll}
0, & L_0 \sqrt{\frac{N T_{\rm obs}}{T_{\rm obs} - (N-1) t_{\rm slew}}}> L_{\rm max}, \\
  \eta \left( L_0 \sqrt{\frac{N T_{\rm obs}}{T_{\rm obs} - (N-1) t_{\rm slew}}}  \right)  , & L_{\rm min} < L_0 \sqrt{\frac{N T_{\rm obs}}{T_{\rm obs} - (N-1) t_{\rm slew}}} < L_{\rm max}, \\
1 , & L_0 \sqrt{\frac{N T_{\rm obs}}{T_{\rm obs} - (N-1) t_{\rm slew}}} < L_{\rm min}. \\
\end{array}
\right. \nonumber \\ \label{RateTargetNfields}
\end{eqnarray}

\begin{figure*}[htbp!]
\begin{center}
\epsfig{file=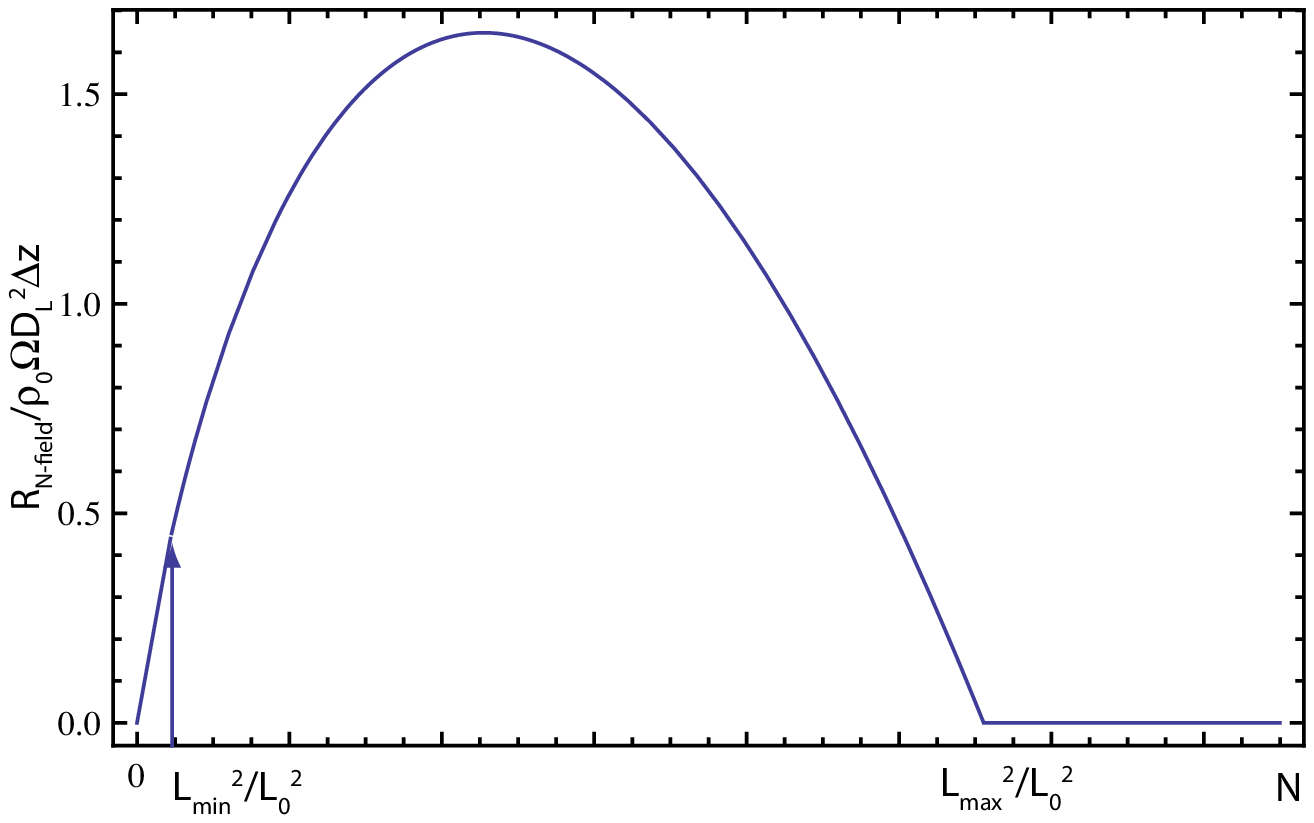, scale=0.7} \\
\epsfig{file=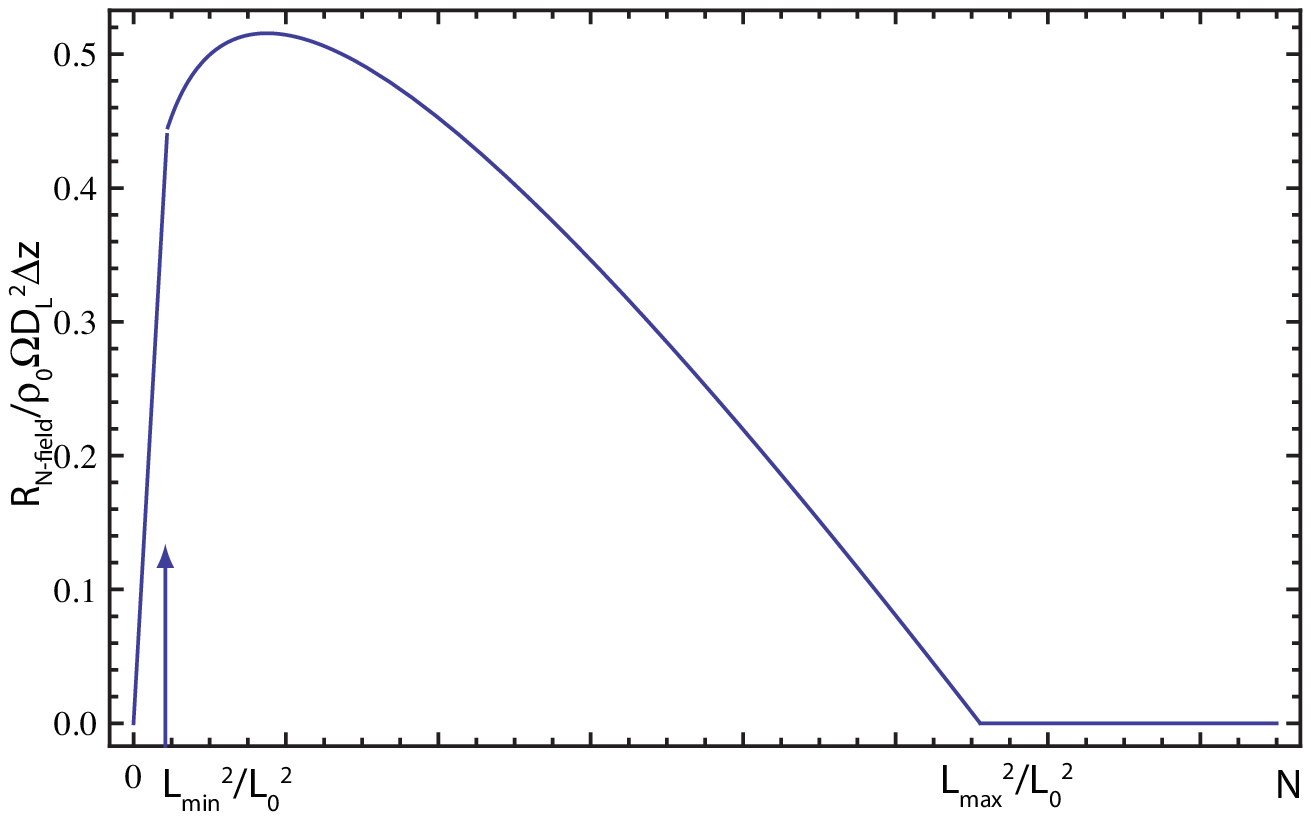, scale=0.7} \\
\epsfig{file=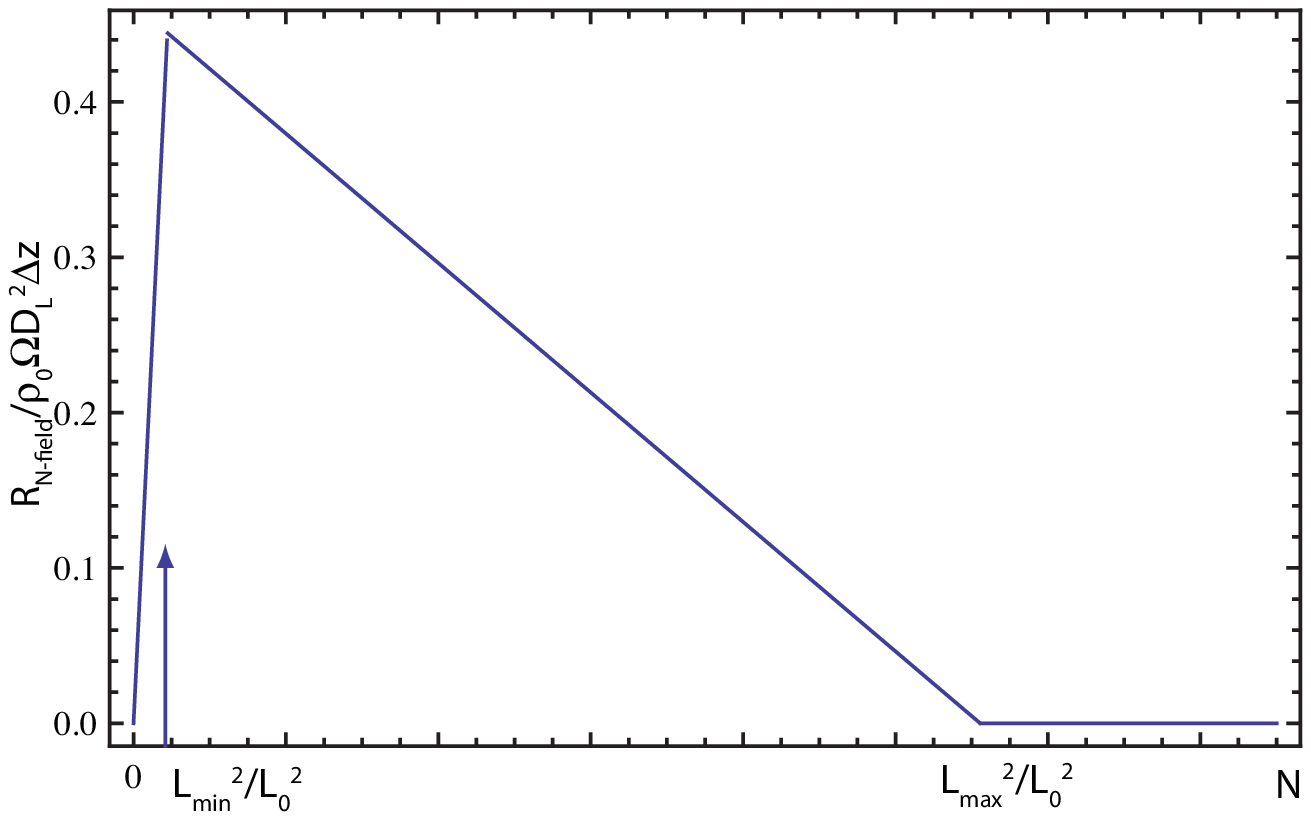, scale=0.7} \\
\end{center}
\caption{The dependence of detection rate on the number of fields surveyed during the interval $T_{\rm obs}$.  These particular plots show the behaviour of detection rate for (top) $\alpha=0.5$, (centre) $\alpha=2.5$ and (bottom) $\alpha =3$. In the simple case shown here in which $t_{\rm slew}=0$, the break points of the rate curve occur at $(L_{\rm min}/L_0)^2$ and $(L_{\rm max}/L_0)^2$.  Note that the peak detection rate occurs in the range $L_{\rm min}^2/L_0^2 < N < L_{\rm max}/L_0^2$ for $0 < \alpha <3$ but occurs at $N < L_{\rm min}^2/L_0^2$ for $\alpha > 3$.} \label{RNfieldPlot}
\end{figure*}

A plot of the behaviour of this function is shown in Figure \ref{RNfieldPlot}. An important characteristic of this function is that the detection rate is found to peak for survey luminosities $L_{\rm min} < L_{\cal N} < L_{\rm max}$ for $0 < \alpha < 3$.  For $\alpha >3$, we see that the optimal detection rate occurs at the point where $L_{\cal N}  = L_{\rm min}$.  The reason for this relates to the steepness of the luminosity distribution.  For distributions steeper than $\alpha >3$, most of the events occur at the low luminosity end, and thus a survey that seeks to maximize detection rate must ensure that it achieves a sensitivity sufficient to reach this minimum luminosity in each field that it surveys.  For such a distribution, the optimal number of fields corresponds to a strategy in which the time available is subdivided so that each field is surveyed to a sensitivity sufficient to detect objects of $L_{\rm min}$ (but obviously no lower).  Conversely, if the luminosity distribution is shallower than $\alpha = 3$, there are not enough events at the low luminosity end to justify spending the integration time on that field to reach the sensitivity to detect them.  The optimal strategy is to increasingly favour field of view over sensitivity as the luminosity distribution becomes progressively shallower.  This point is illustrated in Figure \ref{RNfieldPlot}, where we see that the peak detection rate occurs at larger $N$ for the case $\alpha=0.5$ relative to the case $\alpha=2.5$.

For the case $0< \alpha < 3$, we obtain the location and value of the optimal detection rate by finding the turning point of $R_{\rm N-fields}$.  One has
\begin{eqnarray}
\frac{d R_{\rm N-fields}}{dN} = \left\{ \begin{array}{ll}
\rho_0 \Delta z D_A^2 \Omega, & L_{\cal N} < L_{\rm min}, \\
\rho_0 \Delta z D_A^2 \Omega	
\left[ 
\frac{ L_{\rm max}^{1-\alpha} [T_{\rm obs} - (N-1) t_{\rm slew}] + L_{\cal N}^{1-\alpha} \left[ 
\frac{T_{\rm obs} (\alpha-3)}{2} + t_{\rm slew} \left( N + \frac{\alpha-3}{2} \right)  \right] }{
[T_{\rm obs} - (N-1) t_{\rm slew}] [L_{\rm max}^{1-\alpha} - L_{\rm min}^{1-\alpha}] } \right],
	  & L_{\rm min} < L_{\cal N} < L_{\rm max}, \\
0, & L_{\cal N} > L_{\rm max}. \\ 
\end{array}
\right. 
\end{eqnarray}
For $t_{\rm slew}=0$, the location of the turning point, at $dR/dN=0$, has the simple solution:
\begin{eqnarray}
N_{\rm max} = \left( \frac{L_{\rm max}}{L_0} \right)^2 \left( \frac{3 -\alpha}{2} \right)^{2/(\alpha-1)},
\end{eqnarray}
whose associated maximum detection rate is
\begin{eqnarray}
R_{\rm max} = \Omega \rho_0 \Delta z D_A^2 \left(\frac{1-\alpha}{3-\alpha} \right) 
\left( \frac{L_{\rm max}}{L_0} \right)^2 \frac{L_{\rm max}^{1-\alpha} }{L_{\rm max}^{1-\alpha} - L_{\rm min}^{1-\alpha} } .
\end{eqnarray}
Note that the maximum detection rate depends on $L_0^{-2}$, and thus $S_0^{-2}$.  Thus, in this special instance, the maximum detection rate scales as $\Omega S_0^{-2} \propto \Omega (A_e/T_{\rm sys})^2$, which is the metric employed in some telescope survey figures of merit (Cordes 2007, 2009).

More generally, for $t_{\rm slew} \neq 0$, the maximum detection rate occurs at the value of $N$ which satisfies the following transcendental equation:
\begin{eqnarray}
\left( \frac{ L_{\rm max} }{ L_0 } \right)^{1-\alpha} 
=
\left( \frac{N T_{\rm obs}}{T_{\rm obs} - (N-1) t_{\rm slew}} \right)^{(1-\alpha)/2} 
\left[ \frac{(T_{\rm obs} + t_{\rm slew}) (3 -\alpha) - 2 N t_{\rm slew} }{ 2[ T_{\rm obs} - (N-1) t_{\rm slew}] } \right].
\end{eqnarray}

As a final remark, we note that it is possible in principle for the value of $N_{\rm max}$ to exceed the total number of independent telescope pointings, $N_{\rm point}$, required to cover the entire survey area.  This implies that the optimal strategy includes re-surveying a number of fields for new transients.  A re-observation of a previously visited field can be considered statistically independent of a prior observation, and should detect an entirely new set of transient events, as long as the time between observations of the same field exceeds the event decoherence time (see \S\ref{sec:decoherence} above).  In the opposite limit, in which the event decoherence time is long compared to the interval between observations of the same field, the assumption underpining the foregoing calculation, namely that each observation of a field is statistically independent, is violated.  However, in this case we see trivially that iff $N_{\rm max} > N_{\rm point}$ then the optimal strategy is simply to survey each field only once for a duration $T_{\rm obs}/N_{\rm point}$ (hence $N_{\rm max} = N_{\rm point}$).


\section{Discussion} \label{Discussion}

We have derived in the foregoing section a set of conditions for optimising the tradeoff between survey sensitivity and total field of view.  This enables us to draw some general inferences regarding the optimization of survey strategy.

Although, in a blind transients survey for unknown transient populations, the particulars of the luminosity distribution of events are unknown, we will often be able to place reasonable prior bounds on the luminosity distribution function (i.e. $\alpha$, $L_{\rm min}$ and $L_{\rm max}$ are known only to within a certain range).  To the extent that it is possible to prescribe these bounds it is also possible to specify the range of optimal survey strategies.  Moreover, even when this range is large, the foregoing formalism is useful in determining the optimal means of targetting specific regions of survey parameter space.

We find that, for both surveys of targets at a single fixed distance and surveys over a population of homogeneously-distributed events, the prime consideration in maximizing the detection rate relates to the steepness of the luminosity distribution function.  For distribution functions steeper than $\alpha =3$, the number of events at low luminosity merits an approach which surveys each field down to a sensitivity limit capable of detecting objects at the lowest end of the distribution.  On the other hand, for shallower distributions the highest event rate is achieved by shifting the balance towards greater field of view at the expense of sensitivity.  For $\alpha < 3$ it is advantageous to instead survey events down to an intermediate sensitivity, $S_0 > L_{\rm min}/(4 \pi D_{\rm min}^2)$, and to survey a greater number of fields.  

The dependence on the slope of the luminosity function for a targetted survey can be understood as follows.  The event detection rate is the result of interplay between the increase in the field of view, with which the number of events detected scales linearly, and the increase in sensitivity with time and the associated number of events detectable down to that sensitivity.  The number of objects detectable per field scales as $S_0^{1-\alpha}$ (provided the survey has sufficient sensitivity to be in the regime $S_0 < L_{\rm max}/(4 \pi D_{\rm min}^2)$), and the time required to reach this sensitivity scales as inverse square root of integration time, so the number of objects detected after an integration time $\tau$ on a single field is proportional to a term that scales as $\tau^{(\alpha-1)/2}-C$, where $C$ is a constant.  Now, if we subdivide the time into observations over $N$ separate fields, we see that there is a linear increase in event rate with the number of fields surveyed, but the number of events per field scales as $N^{(1-\alpha)/2}$.  Thus we see that when the index of the luminosity distribution is steeper than $\alpha=3$, the product of the rate per field and the number of fields is a declining function of $N$, and the optimal strategy is to keep $N$ as small as possible by surveying a single field.  On the other hand, for $\alpha<3$ the overall event rate is an increasing function of $N$, at least for small $N$, and thus the optimal strategy is to survey multiple fields of view.

The formalism we have introduced in this paper allows us to address two further facets of survey strategy.  The first relates to the spacing in time of observations of a given survey field, while the second relates to figures of merit for transients surveys.

1. There has been a considerable amount of qualitative discussion within the transients community concerning the most effective means of surveying the variable sky.  A survey which revisits particular field in logarithmically-spaced time intervals is often espoused as optimal on the basis that it detects transients on a variety of timescales (e.g. Murphy et al.\,(2013), section 4).
%

However, if one seeks to optimize the event detection rate, the optimal strategy is to visit any given field on an interval no shorter than the event decoherence time so that no time is wasted redetecting all the slowest-timescale transients in the same field on multiple occasions.  Although the survey will have missed a number of shorter timescale transients in the same field in the meantime, it will have instead detected other short timescale transients while surveying other fields in the intervening time.  

An alternative argument might suggest that the logarithmic sampling approach is advantageous because it performs the dual role of detection and timescale characterization simultaneously.  However, it is difficult to define a metric which quantifies the relative value of both detection and timescale characterization, because it forces one to identify the relative importance of these two dissimilar observational tasks.  There are many other observables (e.g. spectral index, brightness, polarization) by which a transient may be characterised, and these may  obviate the need for timescale characterization in certain circumstances.

2. Some assessments of the efficacy of transients surveys are based on a survey figure of merit which scales as $M_1=\Omega S_0^{-2}$ (Cordes 2007, 2009).  This metric is sometimes adopted as the standard for assessing the survey capabilities of next generation widefield telescopes, and is employed to compare telescope performance in the SKA baseline design.  This metric is appropriate for steady emission (e.g. for steady continuum source and HI surveys) and in quantifying the rate which a given telescope design can survey sky down to a limiting flux density $S_0$. However, {\it the rate at which a transients survey detects events} often exhibits a difference dependence on $S_0$.  It often scales as $M_R=\Omega S_0^{-3/2}$ for surveys of fast transients (see Macquart 2011), and eq.\,(\ref{RateHomog}) indicates that it similarly holds for survey of slow transients in which the events are distributed homogeneously in the survey volume.  For a targetted survey, the event rate scales as $\Omega S_0^{1-\alpha}$.  

This distinction is important, because the metric based on event detection rate indicates a different relative importance between field of view and telescope sensitivity to that in common use.  It bears implications for assessment of telescope designs; a telescope designed to survey for transients based on a metric proportional to $\Omega S_0^{-3/2}$ favours field of view over sensitivity more than one based on the metric $\Omega S_0^{-2}$.  

Interestingly, in a targetted survey for events at a fixed distance, we find that the {\it optimal} detection rate does scale as $S_0^{-2}$.  This result therefore provides a basis for connecting to some earlier metrics for surveys for transients (e.g. Cordes 2007).  However, we stress that this result only holds once the survey is optimized.

\section{Conclusions} \label{Conclusion}

It is possible to optimise slow transients surveys by slewing to multiple independent fields of view within a timescale less than the duration of each transient event and thus trade sensitivity against the total survey area.  We consider an optimization specifically for slow events, namely ones whose duration exceeds the telescope dwell time (see \S\ref{sec:decoherence} for a discussion of this point).  At one extreme it may be optimal to integrate deeply on a small number of fields, and at the other extreme it may be optimal to conduct a shallow search across a large number of fields.  The optimal tradeoff particularly depends on the steepness and upper and lower bounds of the luminosity function and the telescope sensitivity, and a summary of our results is as follows.
\begin{itemize}
\item {\it Surveys of homogenous populations:} For a population of events distributed homogeneously throughout space, the optimum event rate occurs when the maximum number of independent fields of view visited is $N=(T_{\rm decoher} + t_{\rm slew})/4 t_{\rm slew}$, where $t_{\rm slew}$ is the time taken to slew from one field to the next.  This yields an enhancement over the single-field event detection rate of a factor of
\begin{eqnarray}
\frac{3^{3/4}}{4} t_{\rm slew}^{-1/4} \frac{\Delta T_{\rm decoher} + t_{\rm slew}}{\Delta T_{\rm decoher}^{3/4}}. \nonumber 
\end{eqnarray}
 
\item {\it Surveys of quasi-homogeneous populations:} For most homogeneously distributed populations, there is a minimum sensitivity which the survey must attain in order to detect the weakest event at the closest distance at which the events occur, $D_{\rm min}$, with $D_{\rm min} = 0$ corresponding to a completely homogeneous distribution. The optimal number of fields to survey lies in the range $L_{\rm min}/4 \pi S_0 D_{\rm min}^2  < N^{1/2} < L_{\rm max}/4 \pi S_0 D_{\rm min}^2$ provided that the luminosity distribution function is shallower than $\alpha=3$ (neglecting telescope slewing).  We present an equation to derive the optimal number of fields to survey in eq.\,(\ref{N4solns}), which qualitatively behaves as follows.  
\begin{itemize}
\item Since the number of detections scales linearly with the number of fields of view surveyed, there is a balance between survey sensitivity and the total survey area that depends on the steepness of the luminosity function.  For progressively steeper luminosity distributions, more of the events occur at luminosities near $L_{\rm min}$, which biases the event detection rate in favour of searching a smaller number of fields more deeply.
\item For luminosity distributions steeper than $\alpha >3$ the best strategy is to survey only a single field: the maximum detection rate is attained by achieving the sensitivity required to detect all objects all the way down to $L_{\rm min}$ in each field.  Conversely, for shallow luminosity distribution functions, there is progressively less benefit in surveying each field down to $L_{\rm min}$, and the optimal balance between sensitivity and field shifts in favour of field of view.
\item The assumption of true homogeneity ($D_{\rm min}=0$) biases the survey to favour wider fields of view because the survey optimization in part reflects the fact that events may then occur arbitrarily close to the observer, which are best detected by surveying large fields of view at lower sensitivity.  
\end{itemize}

\item {\it Surveys of targetted systems:}
For a survey of a system (e.g. a galaxy or a cluster) at a fixed distance, and neglecting slewing time, the optimal number of survey fields is given by 
\begin{eqnarray}
N_{\rm max} = \left( \frac{L_{\rm max}}{L_0} \right)^2 \left( \frac{3 -\alpha}{2} \right)^{2/(\alpha-1)}, \label{Nmaxagain}
\end{eqnarray}
for $\alpha< 3$, and the associated detection rate is
\begin{eqnarray}
R_{\rm max} = \Omega \rho_0 \Delta z D_A^2 \left(\frac{1-\alpha}{3-\alpha} \right) 
\left( \frac{L_{\rm max}}{L_0} \right)^2 \frac{L_{\rm max}^{1-\alpha} }{L_{\rm max}^{1-\alpha} - L_{\rm min}^{1-\alpha} } .
\end{eqnarray}
Optimization of the survey strategy yields an enhancement in the event rate by a factor of $\sim (L_{\rm max}/L_0)^2$ compared to the detection rate for a single field (eq.\,\ref{RateTarget})).  However, for very steep luminosity functions, $\alpha>3$, the optimal strategy is one in which occurs at the point where the sensitivity satisfies $ N^{1/2} = L_{\rm min}/4 \pi S_0 D_L^2$.

\item In a survey which cycles between a set of fields it is useful to quantify the extent of overlapping event detections between successive visits to the same field.  We present a formalism for quantifying this, and we examine cases in which the distribution of event durations follows Gaussian and power-law distributions.  Event overlap considerations are important when eq.(\ref{Nmaxagain}) implies $N_{\rm max} > N_{\rm point}$ and $\Delta T_{\rm decoher}$ exceeds the time $T_{\rm obs} N_{\rm points}/N_{\rm max}$; in this case the optimal strategy is instead to set $N_{\rm max} = N_{\rm point}$.  

\end{itemize}


\begin{acknowledgements}
Parts of this research were conducted by the Australian Research Council Centre of Excellence for All-sky Astrophysics (CAASTRO), through project number CE110001020.
\end{acknowledgements}


\begin{appendix}
 
\section{The event decoherence function} \label{ACFApp}

In this Appendix we evaluate the event decoherence function explicitly for a succession of events of durations $\Delta T_i$ with random start times $t_i$. The autocorrelation of $N$ is obtained from eq.\,(\ref{Ndefn}) as follows,
\begin{eqnarray}
\langle N(t+t') N(t) \rangle &=& \frac{1}{\cal T} \int_0^{\cal T} dt N(t+t') N(t) = \frac{1}{\cal T} {\cal F} \langle \tilde N(\omega) \tilde N(\omega) \rangle \nonumber \\
&=& \frac{1}{\cal T} {\cal F} \left\langle  \sum_j  |\tilde f(\omega; \Delta T_j)|^2 + \sum_{j \neq k} \tilde f(\omega; \Delta T_j) \tilde f^*(\omega; \Delta T_j) e^{i \omega (t_j-t_k)} \right\rangle. \nonumber \\ \label{BigAvg}
\end{eqnarray}
where the operators ${\cal F}$ and $\tilde \null$ denote a Fourier transform.  Now if the events are distributed randomly in time and there are no correlations in the occurrence times between separate events, then the second term contributes at most a value that is independent of $t'$, namely a constant\footnote{However, this term might be important if some events are connected.  This might occur, for example, if a black hole disruption event spawns a number of related flares.}.  We see this by performing the average $\langle e^{i \omega (t_i-t_j)} \rangle$ as follows.  If the event times are evenly distributed over the interval $[-{\cal T}/2,{\cal T}/2]$ then the probability distribution of $t_j$ is $p(t_j) = 1/{\cal T}, \, -{\cal T}/2 < t_j < {\cal T}/2$, and one has
\begin{eqnarray}
\left\langle \exp \left[ i \omega (t_j-t_k) \right] \right\rangle &=& \frac{1}{{\cal T}^2} \int_{-{\cal T}/2}^{{\cal T}/2} \int_{-{\cal T}/2}^{{\cal T}/2}  dt_j dt_k  \exp \left[ - i \omega ( t_j - t_k ) \right]  \nonumber \\
&=& \frac{2 - 2 \cos {\cal T} \omega}{{\cal T}^2 \omega}.
\end{eqnarray}
Thus the second term in $\langle N(t+t') N(t) \rangle$ is, after substituting in the expression for $\tilde f(\omega,\Delta t_j)$:
\begin{eqnarray}
 &\null& {\cal F} \sum_{j \neq k} \frac{2 - 2 \cos {\cal T} \omega}{{\cal T}^2 \omega} \frac{ 2 \pi }{2\pi {\cal T}} 
\left[ \frac{i}{\omega} - \frac{i e^{i \Delta t_j \omega} }{\omega} \right] \left[\frac{i}{\omega} - \frac{i  e^{i \Delta t_k \omega} }{\omega} \right]^* \nonumber \\
&=& \frac{1}{6 {\cal T}^3} \sum_{j \neq k} 
	\left[ 6 {\cal T} \Delta T_j \Delta T_k + t'^3 {\rm sgn}(t') - (t'-\Delta T_j)^3 {\rm sgn}(t'-\Delta T_j) \right. \nonumber \\
	&\null& \left. \qquad - (t'+\Delta T_k)^3 {\rm sgn}(t'+\Delta T_k) +(t' - \Delta T_j +\Delta T_k)^3 {\rm sgn}(t'-\Delta T_j + \Delta T_k) 	
	\right] \nonumber \\
&\null& \underset{{\cal T} \rightarrow \infty}{\longrightarrow}  \sum_{j \neq k} \frac{\Delta T_j \Delta T_k} {{\cal T}^2}  = \frac{{\cal N} ({\cal N}-1)}{{\cal T}^2} \langle \Delta T_j \Delta T_k \rangle =  \frac{{\cal N} ({\cal N}-1)}{{\cal T}^2} \langle \Delta T \rangle^2
\end{eqnarray}
where in the second line we have used the fact that ${\cal T} \gg \{ t',\Delta T_j, \Delta T_k \}$ and in the third line we have assumed that the durations $\Delta T_j$ are mutually independent.  
 
In the limit in which ${\cal T} \rightarrow \infty$, the autocorrelation reduces to the simple form, 
\begin{eqnarray}
\langle N(t+t') N(t) \rangle &=& \frac{1}{\cal T} {\cal F} \left\langle  \sum_j^{\cal N}  |\tilde f(\omega; \Delta T_j)|^2 \right\rangle 
+ \frac{{\cal N} ({\cal N}-1)}{{\cal T}^2} \langle \Delta T \rangle^2
\nonumber \\
&=&  \frac{1}{2 {\cal T}} \sum_j^{\cal N} \left( |t+\Delta T_j |  + |t-\Delta T_j  | - 2 |t| \right) + \frac{{\cal N} ({\cal N}-1)}{{\cal T}^2} \langle \Delta T \rangle^2. 
\end{eqnarray}  
As we are interested in the correlations between different events, the quantity of interest here is the autocorrelation in the fluctuations in the number of events, $\delta N(t) = N(t) - \bar{N}$, which is given by
\begin{eqnarray}
\langle \delta N(t+t') \delta N(t) \rangle &=&  \left[ \frac{1}{2 {\cal T}} \sum_j^{\cal N} \left( |t+\Delta T_j |  + |t-\Delta T_j  | - 2 |t| \right) \right] + \frac{{\cal N}({\cal N}-1)\langle \Delta T \rangle^2}{{\cal T}^2} -  \left( \frac{{\cal N} \langle \Delta T \rangle}{\cal T}  \right)^2 \nonumber \\
&=& \left[ \frac{1}{2 {\cal T}} \sum_j^{\cal N} \left( |t+\Delta T_j |  + |t-\Delta T_j  | - 2 |t| \right) \right]  - {\cal O} \left( \frac{{\cal N} \langle \Delta T\rangle^2}{{\cal T}^2} \right).
\label{NacfApp}
\end{eqnarray}  
We henceforth ignore the term ${\cal O}(1/{\cal T}^2)$, since this is much smaller than the first term.
This result may be expressed in the equivalent form
\begin{eqnarray}
\langle \delta N(t+t') \delta N(t) \rangle =  \left[  \frac{1}{\cal T} \sum_j^N  \left[ (t'+\Delta T_j) {\rm H}(-t') {\rm H}(t'+\Delta T_j) + (\Delta T_j - t') {\rm H}(t') {\rm H}(\Delta T_j - t') \right]  \right].
\end{eqnarray}
This expression may be evaluated for various distributions of event durations $\Delta T$.

\section{Generalisation of event rates to curved spacetime} \label{AppendixCosmo}
The calculation in Section 2.1 for a set of homogeneously distributed events is only valid when the spacetime geometry is approximately Euclidean.  If the events occur at cosmological distances, the curvature of spacetime needs to be taken into account when computing the event rate.  The foregoing formalism can be modified in a straightforward way to incorporate these effects as follows.  

As before, the maximum distance out to which an object of luminosity $L$ can be detected is
\begin{eqnarray}
D_{\rm max} = \sqrt{ \frac{L}{4\pi S_0}},
\end{eqnarray}
where $D_{\rm max}$ is interpreted as the luminosity distance, which can be related to the redshift of the burst, $z_L$, by the following equation:
\begin{eqnarray}
D_{\rm max}(z_L) = \frac{c (1+z_L)}{H_0} \int_0^{z_L} \frac{dz'}{\sqrt{\Omega_\Lambda + (1-\Omega) (1+z)^2 + \Omega_m (1+z)^3 + \Omega_r (1+z)^4}}.
\end{eqnarray}
For a concordance cosmology one has $\Omega_\Lambda = 0.7$, $\Omega_m=0.3$ and $\Omega=1$ and $\Omega_r=0$.   This luminosity distance corresponds to a comoving distance $D_c = D_L/(1+z_L)$, so the number of events of between luminosities $L$ and $L+dL$ detected down to a limiting flux density $S_0$ is $(4\pi/3) (\Omega/4 \pi) \rho_L dL D_c^3 =  (\Omega/3) \rho_L dL D_L^3 / (1+z_L)^3$.

The event rate calculations now proceed the same as before, and we obtain the following generalization to equation (\ref{RateHomog}):
\begin{eqnarray}
R_{\rm total} = \frac{\Omega \rho_0}{3K} \int_{L_{\rm min}}^{L_{\rm max}} \frac{L^{-\alpha}}{(1+z_L)^3}  \left( \frac{L}{4 \pi S_0 } \right)^{3/2} dL, \label{RtotGen}
\end{eqnarray}
where $z_L$ is the redshift that satisfies the equation $D_{\rm max}(z_L) = \sqrt{L/4 \pi S_0}$. 
We note that our detection rate problem is encountered in other contexts in astrophysics: the problem identical to that encountered when considering number counts of quasars in flux-limited surveys (see, e.g., Longair 1966; Wall 1980; De Zotti et al. 2010).  This event rate reduces to eq.(\ref{RateHomog}) when $z_L \ll 1$, however for $z_L \sim 1$ cosmological effects become non-negligible.  In this case evaluation of the integral in eq.(\ref{RtotGen}) is complicated by the fact that $z_L$ is a function of $L$ but cannot be expressed in terms of $L$ in closed form.  It therefore requires numerical evaluation of $z_L$ as a function of $L$.  

We might hope to gain some insight into the dependence of $R_{\rm total}$ on $S_0$ for cosmological events using various approximations that relate $z_L$ to $L$.  To lowest order in $z_L$, one has $z_L = (H_0/c) \sqrt{L/4 \pi S_0}$, so 
\begin{eqnarray}
R_{\rm total} = \frac{\Omega \rho_0}{3K} \int_{L_{\rm min}}^{L_{\rm max}} \frac{L^{-\alpha}}{\left( 1+ \frac{H_0}{c} \sqrt{\frac{L}{4 \pi S_0}} \right)^3}  \left( \frac{L}{4 \pi S_0 } \right)^{3/2} dL.
\end{eqnarray}
The analytical solution of this integral involves the hypergeometric function $\null_2 F_1$, and it does not yield any immediate further insight into the behaviour of the detection rate.  Further simplification is possible if one takes $(H_0 /c) \sqrt{L/4 \pi S_0} \ll 1$, for which,
\begin{eqnarray}
R_{\rm total} &\approx& \frac{\Omega \rho_0}{3K} \int_{L_{\rm min}}^{L_{\rm max}} \frac{L^{-\alpha}}{\left( 1+ 3 \frac{H_0}{c} \sqrt{\frac{L}{4 \pi S_0}} \right)}  \left( \frac{L}{4 \pi S_0 } \right)^{3/2} dL \nonumber \\
&=& \frac{c \, \Omega \rho_0}{36 \pi H_0  S_0 K (2-\alpha)} \left[ 
L_{\rm max}^{2-\alpha} \,\, \null_2 F_1 \left(1, 2 (\alpha-2); 2 \alpha - 3 ; - \frac{c}{3 H_0} \sqrt{\frac{4 \pi S_0}{L_{\rm max}}} \right) \right. \nonumber \\
&\null& \left. \qquad \qquad \qquad \qquad  
- L_{\rm min}^{2-\alpha} \,\, \null_2 F_1 \left(1, 2 (\alpha-2);2 \alpha -3 ; - \frac{c}{3 H_0} \sqrt{\frac{4 \pi S_0}{L_{\rm min}}} \right)  \right] .
\end{eqnarray}

\end{appendix}
%





\end{document}